\documentclass[aps,prb,
preprint,superscriptaddress]{revtex4-1}
\usepackage{graphicx,subfigure}
\usepackage{dcolumn}
\usepackage{bm}

\begin{document}

\title{
Rod-Like Virus Based Multiarm Colloidal Molecules}

\author{Alexis de la Cotte$^{1,\ddag}$, Cheng Wu$^{1,\ddag}$, Marie Tr\'{e}visan$^1$, Andrii Repula$^1$, Eric Grelet}
 \email{grelet@crpp-bordeaux.cnrs.fr \\
 $^\ddag$ These authors contributed equally to this work}
 \affiliation{Centre de Recherche Paul-Pascal, CNRS \& Universit\'{e} de Bordeaux, 115 Avenue Schweitzer, 33600 Pessac, France}


\begin{abstract}
We report on the construction of multiarm colloidal molecules by tip-linking filamentous bacteriophages, functionalized either by biological engineering or chemical conjugation. 
The affinity for streptavidin of a genetically modified vector phage displaying Strep-tags fused to one 
end of the viral particle, is measured by determining the dissociation constant, $K_{d}$. In order to both improve the colloidal stability and the 
efficiency of the self-assembly process, 
a biotinylation protocol having a chemical yield higher than 90\% is presented to regio-selectively functionalize the cystein residues located at one end of the bacteriophages.
For both viral systems, a theoretical comparison is performed by developing a quantitative model of the self-assembly and interaction of the modified viruses with streptavidin compounds, 
which accurately accounts for our experimental results.
Multiarm colloidal structures of different valencies are then produced by conjugation of these tip-functionalized viruses with streptavidin activated nanoparticles. 
 We succeed to form  stable virus based colloidal molecules, whose number of arms, called valency, is solely controlled by tuning the molar excess. Thanks to a fluorescent labeling of the viral arms, the dynamics of such systems is also presented in real time by fluorescence microscopy.  
\end{abstract}


\maketitle



The design of structured materials at the mesoscopic scale is a challenge that can be addressed with two main strategies usually referred as top-down and bottom-up approaches.\cite{Rev1,Rev2} While the top-down methods consist in the elaboration of patterns and in their progressive size reduction, the bottom-up approach requires the use of nano- or micro-scaled building blocks for which the self-assembly can be driven by entropic or enthalpic contributions allowing for producing hierarchical architectures.\cite{Frenkel} Specifically, most of reported studies have focused on sphere-based assemblies with specific sizes and/or functionalizations opening the field of so-called ``colloidal molecules''.\cite{vanB,Man,Man2} To go beyond the spherical symmetry and developing more complex architectures, some investigations have been performed using nanorods as elementary building blocks.\cite{Salant,Manna,Huang,Fraden,Solomon}
Mainly two distinct paths have been considered for creating multiarm colloidal molecules: the core-first and the arm-first approaches. In the former case, the core is a multivalent initiator from which the arms can be grown. In this way, multivalent-branched nanocrystals \cite{Manna2} or polymer stars can be obtained by a specific crystal growth or by the polymerization of monomers around the core.\cite{Pol1,Pol2,ER3,ER4,ER5,ER6,ER7} This method allows for a control over the length and/or the width of each constituent but limits the construction of objects to a specific valency and a tailored size.  
The second approach consists in the self-assembly of linear constituents into multivalent star- or flower-like structures. This technique requires an end-to-end assembly of the particles which can be achieved in different ways such as DNA templating,\cite{Dujardin,Pal} amphiphilic interactions,\cite{Nie,ER1,ER2} biorecognition \cite{Salant,Huang,Sweeney,Caswell,Linder} or tip-linking reaction of linear polymer chains.\cite{
Pol3,Pol4}
In this paper, we focus on the end-to-end self-assembly of filamentous viruses and aim to obtain multiarm structures of tunable valency (\textit{i.e.} variable number of arms). Such viral rod-like particles can be seen both as elementary building blocks monodisperse in size and shape, and as versatile scaffolds for biological engineering and chemical functionalization of their coat proteins.\cite{Molek,Ng,Cao} Prior studies have reported the use of bacteriophages, mainly modified by phage display, that self-assemble into chains of various length,\cite{Sweeney,Hess} rings \cite{Nam} or radial structures.\cite{Huang,Fraden} These works are based on the genetic engineering of viruses with mutated tip proteins (minor proteins p3 and p9, corresponding respectively to the two extremities of the phage) onto which specific polypeptidic sequences are fused. The formation of chains or rings is then driven either by the complementarity of the involved polypeptidic sequences \cite{Sweeney} or by the introduction of an external crosslinker.\cite{Nam} Star-like structures were also achieved by using the affinity of biotin for streptavidin \cite{Fraden} or 
by biorecognition of a specific motif Histidine-Proline-Glutamine (HPQ),\cite{Devlin} called Strep-tag, with streptavidin coated nanoparticles.\cite{Huang} Specifically, a mutant called M13-AntiStreptavidin (M13AS), initially isolated through the screening of a phage display library, has been already used in templating hierarchically organized hybrid materials.\cite{SWLee,Huang,SWLee2016} Here, we study this M13AS mutant displaying Strep-tags at one end by first quantifying its affinity for streptavidin with the determination of its dissociation constant in solution, $K_{d}$. 
Because of the relatively high dissociation constant $K_{d}$ found, limiting therefore the self-assembly efficiency for the formation of colloidally stable multiarm structures, we have developed a chemical alternative approach based on the biotinylation of cystein residues present at one tip (p3 proteins) of the so-called M13C7C bacteriophages. The protocol relies on the thiol based bioconjugation, and it has been optimized to be both highly specific (no other coat protein modified) and efficient (yield of about 92\%). We then discuss a quantitative model accounting for the interaction  with streptavidin of both the biologically and the chemically functionalized phages for various initial conditions. 
When exposed to a dispersion of streptavidin coated nanoparticles, biotinylated phages self-assemble into colloidally stable multiarm structures, as the affinity of streptavidin for biotin is several orders of magnitude higher than for Strep-tags. Moreover, these multiarm molecules display a tunable valency which can be continuously monitored by varying the relative molar excess of biotinylated phages and streptavidin activated nanoparticles. This approach makes also possible a second chemical functionalization of the phages by their body labeling with fluorescent dyes, allowing for the \textit{in situ} observation by optical microscopy of their self-assembly into multiarm colloidal molecules.

\section*{Results and discussion}

\subsection*{M13AS Binding Affinity and Biotinylation Yield of M13C7C-B}

\begin{figure}
  \includegraphics[width=0.7\columnwidth]{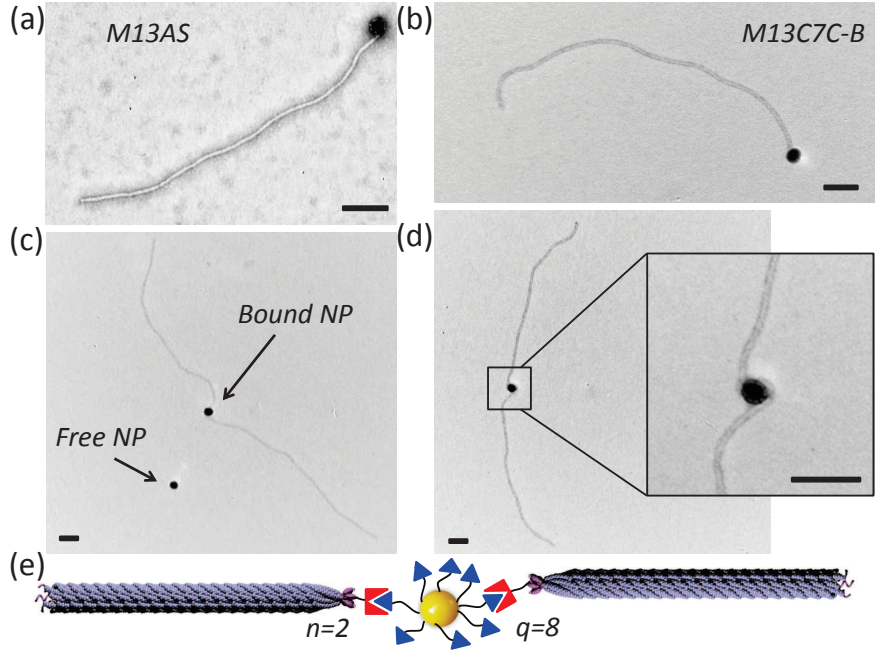}
  \caption{\label{Fig1} Transmission electron microscopy (TEM) images of M13AS ((a) and (c)) and M13C7C-B ((b) and (d)) viruses complexed with streptavidin-coated nanoparticles (NP). (a),(b) Single viruses and (c),(d) pairs of viruses bound to a nanoparticle. The scale bars represent 100~nm. (e) Schematic representation of two viruses ($n=2$) bound to a NP exhibiting $q=$8 streptavidin proteins at its surface.}
\end{figure}

In our mixtures of M13 viruses (either AntiStreptavidin phages, M13AS or biotinylated phages, M13C7C-B) and streptavidin coated nanoparticles (NP) exhibiting $q$ streptavidin molecules (Strep) per NP, the principle of mass conservation applied on virus and streptavidin species provides the two following relations:

\begin{equation}
[M13]_{0} =
[M13]_{free} + [M13]_{bound}
\label{eq:M13ASmass}
\end{equation}

with [M13]$_{0}$ the initial virus concentration, [M13]$_{free}$ and [M13]$_{bound}$ the respective concentrations at equilibrium of free and bound (or reacted) viruses with streptavidin coated nanoparticles;

\begin{equation}
q\times [NP]_{0} =
\left
(q \times [NP]_{free} + (q-n)\times [NP]_{bound}
\right) + n\times [NP]_{bound}
\label{eq:NPmass}
\end{equation}

where $n$ is the number of viruses bound per NP ($1\leq n \leq q$), [NP]$_{0}$ and [NP]$_{free}$ the respective initial and unreacted concentrations of nanoparticles, and [NP]$_{bound}$ the concentration of nanoparticles conjugated with at least one M13 virus (Fig. \ref{Fig1}). The binding affinity of a ligand for a receptor is given by the dissociation constant, $K_{d}$, which results from the following equilibrium in solution:\cite{Sanders}

\begin{equation}
M13_{free} + Strep_{free}  \rightleftharpoons M13_{bound} 
\nonumber
\end{equation}

and is defined as:

\begin{equation}
K_{d} \equiv
\frac{[M13]_{free} \times [Strep]_{free}}{[M13]_{bound}}
\label{eq:Kd}
\end{equation}

 where [Strep]$_{free}=q \times [NP]_{free} + (q-n)\times [NP]_{bound}$. Qualitatively, the lower the $K_{d}$ value is, the more stable the complex is and thus the stronger the affinity.
 In the case of the interaction between biotin and streptavidin protein, 
 $K_{d}$ reaches a value of 10$^{-15}$~M, considered the strongest non-covalent bond found in nature.\cite{Thermo}
 The Histidine-Proline-Glutamine (HPQ) motif is a well-known Strep-tag with however a lower affinity for streptavidin than biotin, typically found in the $\mu$M range.\cite{Devlin} However, the affinity of M13AS phage where the HPQ motif is displayed on each of its five p3 proteins, has not yet been determined.
 Using Eqs. \ref{eq:M13ASmass} and \ref{eq:NPmass}, let's rewrite $K_{d}$ as a function of an unique variable, the fraction of bound nanoparticles $f(NP)\equiv [NP]_{bound}/[NP]_{0}$:
 
 \begin{equation}
 K_{d} = 
 \left(
 \frac{[M13]_{0}}{n \times f(NP)} - [NP]_0
 \right)
 \times 
 \left(
 q-n \times f(NP) 
 \right)
 \label{eq:KdNP}
 \end{equation}
 
\begin{figure}
	\includegraphics[width=1\columnwidth]{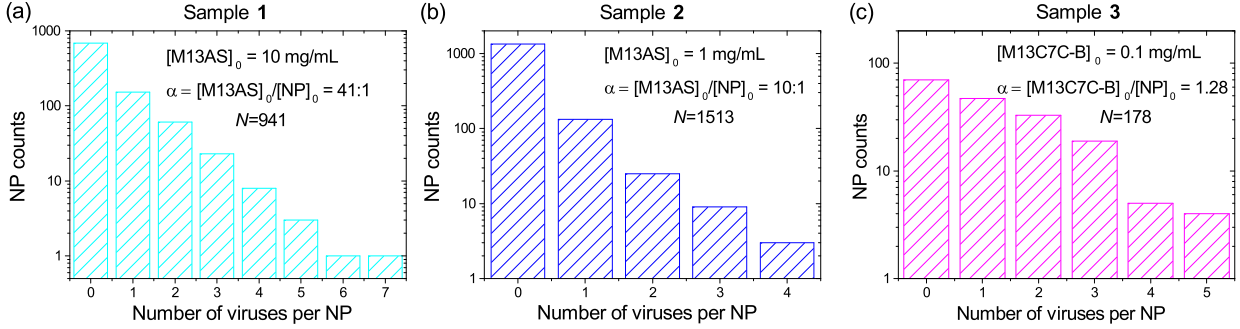}
	\caption{\label{Fig-counts} Distribution of the number of bound viruses per 
	nanoparticle in suspensions of (a) M13AS at initial concentration of 10 mg/mL (5.4  $\times 10^{-7}$~M) with a molar excess of virus per nanoparticle $\alpha=$~41 (Sample \textbf{1}), (b) [M13AS]$_0$=1 mg/mL (5.4 $\times 10^{-8}$~M) with $\alpha=$~10 (Sample \textbf{2}), (c) biotinylated viruses at [M13C7C-B]$_0$=0.1 mg/mL (5.4 $\times 10^{-9}$~M) with $\alpha$=1.28 (Sample \textbf{3}). $N$ indicates the total number of NP counted by TEM for establishing the statistics of each sample. }
\end{figure}

 Thanks to Eq. \ref{eq:KdNP}, the M13AS-streptavidin dissociation constant, $K_{d}$, can be determined by counting by TEM both $f(NP)$ and $n$, the average number of viruses per nanoparticle having at least one virus bound (defined also as the valency of the self-assembled structure). Note that using this model, a precise estimation of $K_{d}$ can only be performed if no steric hindrance limits the binding of viruses to nanoparticles. This means practically that $n \ll q$, 
 as experimentally confirmed in Fig. \ref{Fig-counts}(a) and (b) for which $n\simeq~1.5$ (see Table \ref{tab:NP}). 
 
As M13AS viruses are in large excess compared to NP (see Table \ref{tab:NP}), only the concentrations at equilibrium of free and bound NP as well as the average number of viruses grafted per NP, $n$, can been counted by TEM on a given sample ($[M13AS]_{free}$ being too high to be determined at the same time). 
In order to determine $K_{d}$,  
two specific mixtures of M13AS and streptavidin-coated nanoparticles (Samples \textbf{1} and \textbf{2}, described in Table \ref{tab:NP}) were realized with different initial conditions, for which the molar excess is introduced, $\alpha$:


\begin{equation}
\alpha \equiv
\frac{[M13]_{0}}{[NP]_{0}}
\label{eq:alpha}
\end{equation}

\begin{table}
\caption{\label{tab:NP} Summary of the experimental conditions and measurements obtained by TEM for the three mixtures of viruses and streptavidin coated nanoparticles. Samples \textbf{1} and \textbf{2} correspond to biologically engineered M13AS phages and sample \textbf{3} to tip-biotinylated M13C7C-B viruses. The fraction of bound nanoparticles, $f(NP)$, and the average number of viruses per bead, $n$, result from TEM counting shown in the distributions plotted in Fig. \ref{Fig-counts}. The dissociation constant, $K_{d}$, is calculated for samples \textbf{1} and \textbf{2} according to Eq. \ref{eq:KdNP}, and is taken from literature \cite{Thermo} for sample \textbf{3}.}
{\setlength{\extrarowheight}{10pt}
\begin{tabular}
	{c c c c}

  \multicolumn{1}{c}
  {Sample} & \textbf{1} & \textbf{2} & \textbf{3} \\
  \hline \hline 
  Virus & M13AS & M13AS & M13C7C-B \\
 
  $[M13]_{0}$~(M)& 5.4~$\times$~10$^{-7}$ & 5.4~$\times$~10$^{-8}$ & 5.4~$\times$~10$^{-9}$ \\
    
  $[NP]_{0}$~(M)& 1.3~$\times$~10$^{-8}$ & 5.4~$\times$~10$^{-9}$ &  4.2~$\times$~10$^{-9}$ \\
   
  $\alpha=[M13]_{0}/[NP]_{0}$& 41 & 10 & 1.28\\
  
  $f(NP)$ & 0.27 & 0.11 & 0.61\\
   
  $n$ & 1.62 & 1.31 & 1.94 \\
  
  
   
  $K_{d}$~(M)& 9~$\times$~10$^{-6}$ & 3~$\times$~10$^{-6}$ & 10$^{-15}$ \\
 \hline
\end{tabular}}
\end{table}

Both mixtures were observed in TEM (Fig. \ref{Fig1}) and the distributions of the number of viruses grafted per bead is provided in Fig. \ref{Fig-counts}(a) and (b) for samples \textbf{1} and \textbf{2}, respectively. The mean values of $n$ and $f(NP)$ are deduced from these distributions, and consequently 
$K_{d}$ is obtained thanks to Eq. \ref{eq:KdNP}. A similar dissociation constant (within the experimental margin of error) $K_{d}$=6~($\pm$~3)~$\times$~10$^{-6}$~M is found for both samples (Table \ref{tab:NP}). This average value is also in good agreement with the one reported in literature for the Histidine-Proline-Glutamine Strep-tag of about 1~$\mu$M.\cite{Devlin} The slightly lower binding affinity found in our case can be explained by the following reason: The determination of binding affinity is usually performed with the target compound (here: streptavidin) deposited on a solid surface, whereas our experiments were performed in bulk with dispersions of streptavidin coated NPs. These NPs not only diffuse slower than molecular streptavidin, but also induce steric hindrance as soon as some viruses are bound making more difficult further binding of other viruses.  
Nevertheless, this relative high value of the dissociation constant, for which $K_d\gg[M13AS]_0\geq[NP]_0$ (Table \ref{tab:NP}), does not allow for a tight and non-reversible binding. A first alternative would be to design
a Strep-tag with an affinity in the nM range, providing then a suitable efficiency for high enough initial virus ($>$~1mg/mL) and bead concentrations. The second alternative is to work with biotin derivatives ($K_d\simeq 10^{-15}$~M), which have been chemically grafted at one of the virus tips (see \textit{Materials and methods}). In order to determine the biotinylation yield of the M13C7C-B phages, sample \textbf{3} has been prepared (Table \ref{tab:NP}), where phages and streptavidin activated nanoparticles have been introduced in similar amount ($\alpha\simeq 1$,) allowing for their simultaneous counting by TEM (Fig. \ref{Fig1}). Specifically, among a total number of 229 counted viruses, 210 were bound by their tip to nanoparticles and 19 were found free, giving a fraction of bound M13C7C-B of $f(M13)=0.92$ and therefore a yield of biotinylation of $92\%$. It is worth mentioning that the functionalization of the virus tips by biotin is not only efficient, but also highly selective as shown by the absence of any virus based by-products. Moreover, the specificity of the reaction between beads and functionalized phages has been checked by a control experiment with raw M13 viruses, and, as expected no interaction is observed with nanoparticles (see Figure S3 in Supporting Information).  

\subsection*{Quantitative Model for the Formation of Virus/NP Structures}

\begin{figure}[!h]
	\includegraphics[width=0.5\columnwidth]{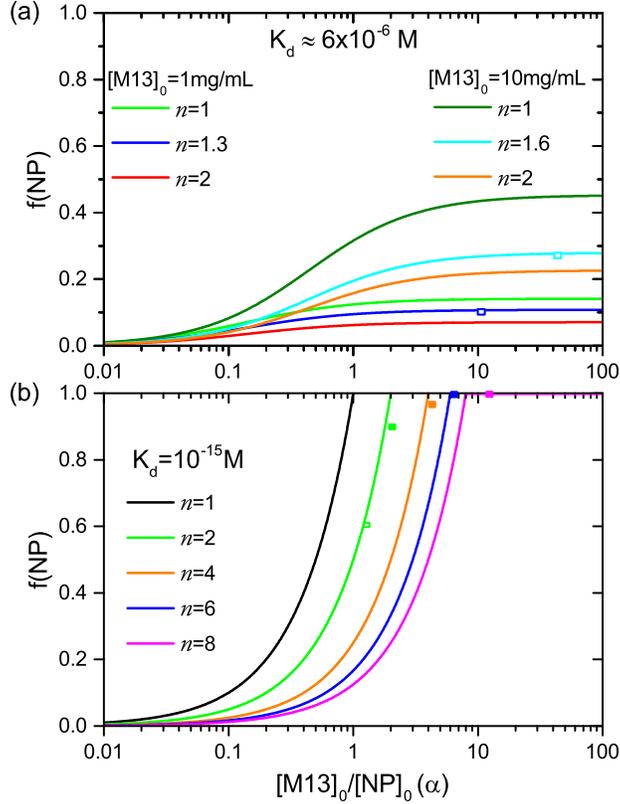}\\
	\caption{\label{Fig:Model} Fraction of bound nanoparticles, $f(NP)$, as modelized by Eq. \ref{eq:model}, as a function of the molar excess, $\alpha$, obtained for different average numbers of viruses per NP, $n$. Results calculated for: (a) M13AS virus strain with K$_{d}\simeq6\times 10^{-6}$~M for two different initial virus concentrations, [M13AS]$_{0}$;  
		(b) biotinylated viruses, M13C7C-B, with K$_{d}= 10^{-15}$~M. In this case, the results are independent of the initial virus concentration, as [M13AS]$_{0}\gg $K$_{d}$, and only depend on the molar excess $\alpha$. The cyan, blue and green open square symbols represent the samples \textbf{1}, \textbf{2} and \textbf{3}, respectively, as described in Table \ref{tab:NP}. The green, orange, blue and pink full square symbols represent the four experimental samples of respective average valency $n=2.1, 4.2, 6.1$ and 8.3, as reported in the next Section. }
\end{figure}
Rewriting the expression of $K_{d}$ given in Eq. \ref{eq:KdNP}, the fraction of bound nanoparticles $f(NP)$ can be accounted by the following quadratic equation:

\begin{equation}
{f(NP)}^2 - \frac{1}{n}
\left(
\frac{\alpha K_d }{[M13]_0}  + q + \alpha
\right)f(NP) + \frac{q \alpha}{n^2}
=0
\label{eq:model}
\end{equation}

Eq. \ref{eq:model} can be easily solved analytically to provide $f(NP)$ for different initial conditions. 
The results are plotted in Fig. \ref{Fig:Model}. When the binding affinity for streptavidin is low, as for M13AS viruses, the self-assembly yield to form multiarm structures ($n\geq2$) remains weak ($f(NP)<25\%$), whatever the initial virus and NP concentrations (Fig. \ref{Fig:Model}(a)). At a fixed value of the dissociation constant $K_{d}$, only the initial conditions affect the equilibrium, as illustrated by the two experimental data points corresponding to samples \textbf{1} and \textbf{2}, which are in very good agreement with the predictions of the model.
Fig. \ref{Fig:Model}(b) shows that strongly increasing the binding affinity favors the formation of multiarm structures, for which a total reaction can be obtained. More importantly, the valency of multiarm self-assemblies can be controlled and tuned \textit{only} by varying the molar excess $\alpha$. Finally, an outstanding agreement between our model and sample \textbf{3} is found  (Fig. \ref{Fig:Model}(b)), especially considering the absence of any free parameter. In conclusion, our model quantitatively accounts for the self-assembly behavior of our two experimental systems, and it confirms that biotinylated phages are the most suitable one to form multiarm colloidal molecules, as described in the next paragraph. 

\subsection*{Construction of Multiarm Colloidal Molecules}


\begin{figure}
	\includegraphics[width=1\columnwidth]{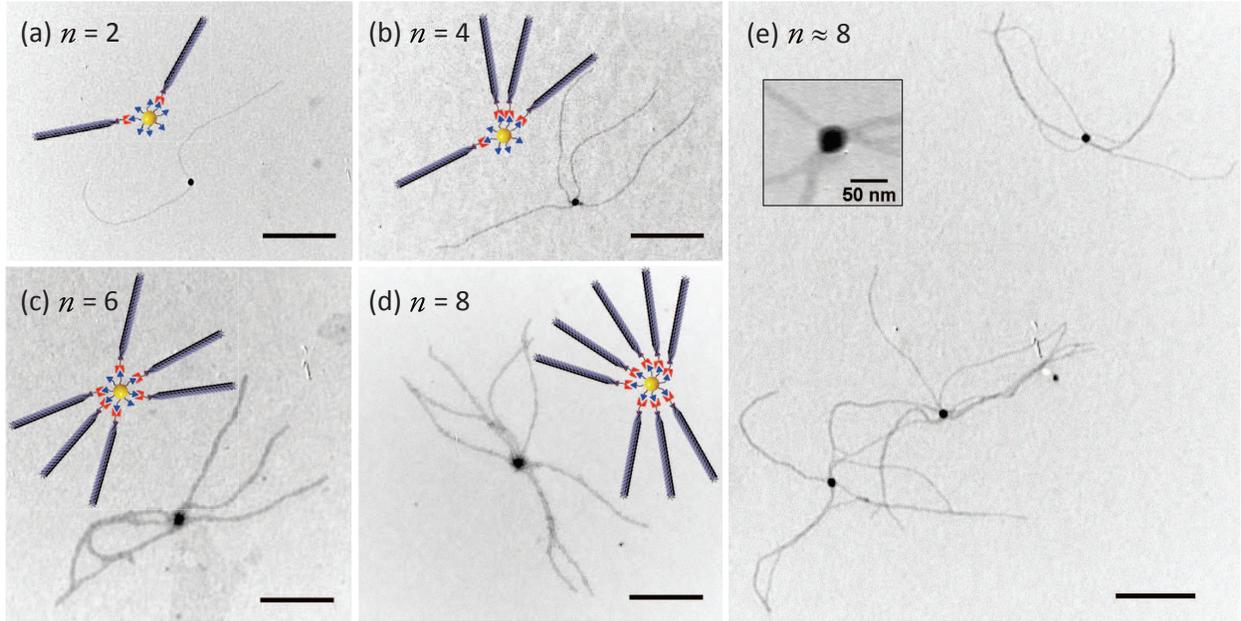}\\
	\caption{\label{Fig:M13C7CBS} Virus based multiarm self-assemblies observed by TEM obtained from the reaction between tip-biotinylated phages (M13C7C-B) and streptavidin coated nanoparticles.
		The number of arms can be continuously tuned, as illustrated with structures of valency $n=$2 ((a), dimer), 4 ((b), tetramer), 6 ((c), hexamer) and 8 ((d), octamer). (e) Large field of view with three multiarm molecules of average valency $n=$8. Inset: zoom-in of the multiarm structure core. Except in the inset, the scale bars represent 500~nm.}
\end{figure}

\begin{figure}
	\includegraphics[width=0.9\columnwidth]{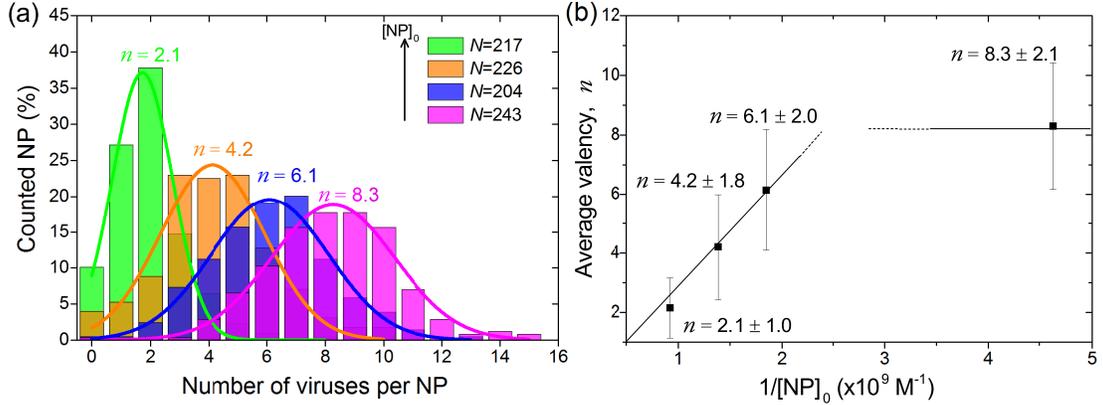}\\
	\caption{\label{Fig:M13C7CBSStat} (a) Distribution of the number of viruses per nanoparticle for four different samples, at fixed biotinylated virus concentration, [M13C7C-B]$_0$, and increasing nanoparticle concentration, [NP]$_0$. The arithmetic mean, or average valency $n$, is indicated and the solid lines correspond to Gaussian fits. $N$ is the total number of NP counted by TEM for establishing the statistics of each sample. (b) The average valency $n$ increases linearly with the molar excess, and saturates when $n$ reaches the maximum number of available streptavidin proteins per NP, $q \simeq 8$. The error bars are the standard deviation obtained from the Gaussian fits. 
	}
\end{figure}

\begin{figure}
	\includegraphics[width=0.6\columnwidth]{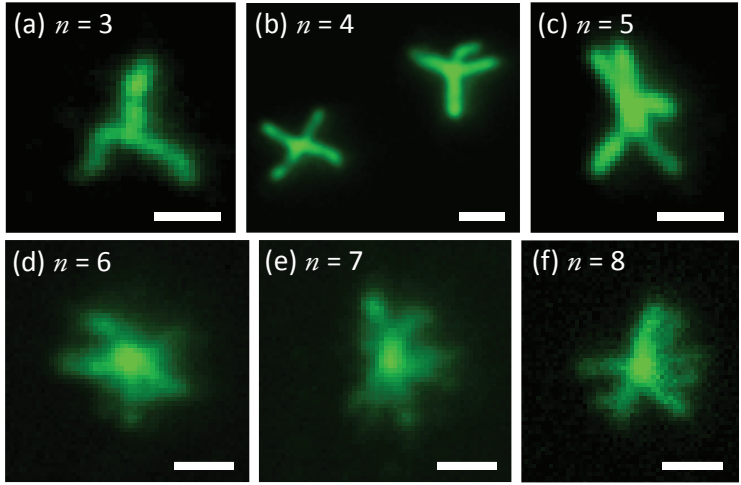}\\
	\caption{\label{Fig:M13C7CFluo} Optical fluorescence microscopy images showing multiarm colloidal molecules formed by fluorescently labeled M13C7C-B phages. Different valencies, $n$, can be observed \textit{in situ} in aqueous dispersions. 
		The scale bar, identical for the six images, represents 1~$\mu m$.}
\end{figure}

Multiarm colloidal molecules were achieved by preparing four samples of M13C7C-B phages mixed with aqueous dispersions of streptavidin coated nanoparticles (see \textit{Materials and methods} section). As shown in Fig. \ref{Fig:M13C7CBS} and in Supplementary Fig. 4, the rod-based colloidal molecules are radial structures possessing a single central NP surrounded by up to $n=8$ arms formed by tip-linked M13C7C-B functionalized phages. 
Figure \ref{Fig:M13C7CBSStat} displays the increase of the average valency $n$ of multiarm assemblies when the initial concentration of NP
is decreased at a fixed virus concentration. 
This behavior is \textit{quantitatively} consistent with our theoretical model, as plotted in Fig. \ref{Fig:Model}(b) where the full symbols represent the four samples of colloidal molecules reported in Fig. \ref{Fig:M13C7CBSStat}. We experimentally confirm that the molar excess is the key-parameter to tune the valency of the self-assemblies. After a linear increase, the average valency $n$ saturates in Fig. \ref{Fig:M13C7CBSStat}(b) when it reaches the maximum number of available streptavidin binding sites per bead, $q$. At this saturated value, $n$ becomes independent of the initial molar excess $\alpha$ (for $\alpha \geq q$), as predicted by the model (Fig. \ref{Fig:Model}(b)). 
In the results shown in Fig. \ref{Fig:M13C7CBSStat}(a), all samples present a similar distribution of virus populations (except for the lowest average valency $n=2.1$ which is asymmetric), whose width, defined as the standard deviation of the corresponding Gaussian fits, is $\delta n \pm 2$. This width of the average valency is narrow enough 
to obtain slightly polydisperse (in terms of number of arms) batches of star-like colloidal molecules for the four different prepared samples (corresponding to the four different initial NP concentrations).  
Thanks to an efficient and regio-selective biochemical functionalization of the M13C7C phages followed by a conjugation with streptavidin activated beads, the large scale production of multiarm structures of well-defined and tunable valency has been achieved. Furthermore, biotinylated phages are an interesting system to work with because of their ability to be twice chemically modified without losing their functionality. We have succeeded in labeling with fluorescent dyes the virus body, targeting mainly the p8 major coat proteins. Labeled M13C7C-B phages can therefore be visualized at the single particle level by optical fluorescence microscopy (Fig. \ref{Fig:M13C7CFluo}), allowing for \textit{in situ}, \textit{i.e.} at equilibrium, sample observation as shown in the movie presented in Supporting Information. 
If the structures are identical in shape to those seen by TEM (Fig. 
\ref{Fig:M13C7CBS}), information on the dynamics of the different multiarm structures are here available. Remarkably, some flexibility can be noticed between the arms of the self-assemblies, contrary for instance to multipod structures obtained by crystal growth.\cite{Solomon} This flexibility in the link between bound viruses is distinguishable from the Brownian motion of the multiarm structures themselves. Such \textit{in situ} observations 
provide the experimental equilibrium state of virus based multiarm structures.   

\section*{Conclusions}

In this paper, we have studied the self-assembly of functionalized filamentous viruses into multiarm colloidal structures. First, we used an engineered phage (M13AS), modified at one end to display a polypeptidic sequence (HPQ motif) exhibiting affinity for streptavidin. We determined the dissociation constant of the system in solution to be  $K_{d} = 6 \times$ 10$^{-6}$~M consistent with literature value for this protein tag. 
In order to strengthen the affinity for streptavidin, a selective and efficient biotinylation protocol of one of the phage tips has been successfully performed. When conjugated with streptavidin activated nanoparticles, these chemically modified phages self-assemble into colloidally stable multiarm structures, whose valency has been continuously tuned from 1 to 8 by varying the molar excess. An additional chemical functionalization of the phage body by fluorescent dyes has been done to optically visualize the multiarm colloidal molecules at the single particle scale and to get \textit{in situ} information on their dynamics. 
The results presented here show how regio-functionalized viruses can be used as building blocks for mesostructured materials and for the development of a next generation of rod based colloidal particles.

\section*{Materials and methods}

M13-AntiStreptavidin phage presents Strep-tags on each of its five p3 proteins localized at one tip. M13AS was kindly provided by S.-W. Lee from the University of Berkeley (USA). M13AS was isolated through the screening of a phage display library and several rounds of selection for streptavidin, with the following inserted sequence on the p3 proteins, identified by DNA sequencing (see Figure S1 in Supporting Information for the complete sequencing): AC\textbf{\underline{HPQ}}GPPCGGGS.\cite{Yoo} The specific motif Histidine-Proline-Glutamine (HPQ),\cite{Devlin} called Strep-tag, is known for its affinity with streptavidin. Large amplification of these phages of molecular weight $M_w=1.85\times10^7$~g/mol was performed using standard biological procedures.\cite{Sambrook} Before further use, M13AS phages were extensively dialyzed against PBS buffer (pH 7.5, I=20~mM).   

A second filamentous phage, called M13C7C and displaying cystein residues on the exposed part of each p3 protein (cystein amino acids otherwise absent from the other coat proteins of the phage), has been used in this study and bioconjugated with maleimide activated biotin. Infecting solution was purchased from New England Biolabs (NEB, USA), and corresponds to a mixture of several M13C7C clones. The amino acid sequences inserted on each of the five p3 proteins of the viruses display 7 random amino acids flanked by two oxidized cysteins which form a disulfide bridge. After phage titration, we isolated a plaque consisting of a colony of bacteria infected by a single virus clone and thus a unique amino acid sequence (see Figure S2 in Supporting Information for the complete DNA sequencing). M13C7C phages were also grown and purified using standard biological protocols.\cite{Sambrook}

In order to have free thiols available for conjugation with maleimide activated compounds, a reduction step of the disulfide bridges is required. This was done by introducing a reducing agent, as Tris(2-carboxyethyl)phosphine hydrochloride (TCEP, Thermo Scientific) used at a concentration around 2~mM.\cite{Heinis, Heinis2} To prevent any re-oxidization of the free thiols and the reformation of the disulfide bridges, all the reaction process occurred in degassed media and in the presence of 1 mM of ethylenediaminetetraacetic acid (EDTA, Sigma Aldrich).\cite{Heinis} The viruses were first depleted by the addition of 20~\% (v/v) of a PEG-8000/NaCl mixture (200 g/L of PEG-8000, 2.5 M of NaCl, degassed), then redispersed in 50 mM Tris (+ 1 mM EDTA, degassed) to have a final concentration of about 1 mg/mL.
The dispersion of phages in presence of TCEP was then stirred (400~rpm) at 4$^{{\rm o}}$C during 48 hours.\cite{Ng,Derda2}
After the reduction reaction, viruses were purified by a centrifugation step (13.5~kg during 20 minutes) followed by 6 rounds of 10~min dialysis against PBS buffer (pH 6.5, I=350~mM + 1~mM EDTA, degassed). Maleimide-PEG2-biotin (Thermo Scientific) was dissolved in distilled water (Maximum solubility: 25~mg/mL). A 1000-fold molar excess was taken per virus and added drop-wise to the suspension. The mixture was kept under inert atmosphere thanks to the initial bubbling of Argon, protected from light and stirred during 2 hours at room temperature. 
The resulting phages after biotinylation, called M13C7C-B, were then dialyzed against Bis-Tris (I=20 mM pH 7.4, 0.02\% NaN3) and purified by rinsing with the same buffer in a 100k MWCO filter tube (Amicon Ultra, Merck Millipore).\\ 

The fluorescent labeling of the M13C7C-B was achieved 
in PBS buffer (pH 7.8, I = 350~mM) at a concentration of 0.2~mg/mL. An excess of 3 dyes (either Alexa488-NHS ester activated, Molecular Probes, or Fluorescein-NHS ester) per p8 main coat protein (about 3000 proteins per virus) first dissolved in N,N-dimethyleformamide (DMF) was added to the suspension. Particular care was taken that the proportion of DMF does not exceed 20\% of the total volume in order to avoid any denaturation of the phages by the organic solvent. The mixture was protected from light and kept under stirring at room temperature during 2~hours. The resulting viruses 
were then extensively dialyzed against Bis-Tris buffer (pH 7.4, I=20~mM).

To construct multiarm structures with M13C7C-B as well as to estimate the binding affinity of M13AS with streptavidin, the viruses were mixed with 30~nm streptavidin coated iron oxide nanoparticles (Ocean NanoTech) for 14 hours at room temperature. The number of available streptavidin proteins per bead for conjugation with phages, $q$, is estimated to be of about 8 (maximum mean valency observed). Rather large nanoparticles have been chosen in order to make easier their counting by transmission electron microscopy (TEM) experiments. In particular, non-specific interactions between viruses can then be easily discriminated, especially when compared with the use of free molecular streptavidin whose relative small size makes tricky its observation by TEM. Note that some streptavidin release occurred with time due to nanoparticle aging (life time of 3 months).      

Samples were then observed by TEM on a Hitachi H-600 microscope operating at 75~kV and images were recorded with an AMT CCD camera. The diluted virus suspension (typically 10$^{-2}$ - 10$^{-3}$~mg/mL) was settled onto freshly treated (O$_2$ plasma, K1050X, Quorum Technologies) 
200-mesh formvar/carbon-coated grids (Agar) and stained with 2\% (w/w) uranyl acetate.

The fluorescent labeled virus particles were visualized using an inverted microscope (IX-71, Olympus) equipped with high numerical aperture (NA) oil objective (100x PlanAPO NA 1.40, Olympus) and a Neo sCMOS camera (Andor).

\section*{Associated content}

Supporting Information: DNA sequencing of the two bacteriophages, large field of view images by TEM of the control experiment with raw viruses and of the multiarm structures, and movie of multiarm colloidal molecules observed by fluorescence microscopy.

\section*{Author Contributions} 
E.G. designed the project and supervised the research. C.W. and A.C. prepared the biotinylated phages and studied their self-assembly, M.T. prepared the samples and performed the experiments on M13AS viruses, C.W. and A.R. performed the fluorescence microscopy experiments. E.G. and A.C. developed the theoretical model, and wrote the manuscript.

\begin{acknowledgments}

This research was supported by the French National Research Agency (ANR) through the project AURORE. We thank
Pr S.-W Lee 
for the generous gift of M13AS and P. van der Schoot for useful discussion. A.R. acknowledges support from the European Union's Horizon 2020 research and innovation programme under the Marie Sklodowska-Curie Grant Agreement No 641839.

\end{acknowledgments}

\end{document}